# MEDEAS-World model calibration for the study of the energy transition

AUTHORS:
Gianluca Martelloni
Francesca Di Patti
Ilaria Perissi
Sara Falsini
Ugo Bardi
Corresponding author: Gianluca Martelloni (email address: gianluca.martelloni@unifi.it)

**ABSTRACT** MEDEAS ("Modelling the Energy Development under Environmental And Socioeconomic constraint") World is a new global-aggregated energy-economy-environmental model, which runs from 1995 to 2050. In this work, we tested the MEDEAS world model to reproduce the IPCC (International Panel on Climate Change) GHG (Green House Gases) emission pathways consistent with 2 °C Global Warming. We achieved parameter optimizations of the MEDEAS model related to different scenarios until 2050. We chose to provide a sensitivity analysis on the parameters that directly influence the emission curves focusing on the annual growth of the RES (Renewable Energy Sources), GDP (Gross Domestic Product) and annual population growth. From such an analysis, it has been possible to infer the large impact of GDP on the emission scenarios.

**INTRODUCTION**

MEDEAS world model is a global, one region-aggregated economy-energy-environment model which has been developed applying System Dynamics to integrate the knowledge from different perspectives as the feedbacks from different subsystems.

MEDEAS world consists of a modular and flexible structure, where each module can be expanded/simplified/replaced by another version or sub-model. MEDEAS model runs from 1995 to 2050 in order to predict the energy transition Fossil Fuels/RES and it is structured into 7 submodules (Capellán-Pérez et al., 2017a; 2017b; Solé et al., 2018):

1) **Economy**: It is modelled following a post-Keynesian approach assuming disequilibrium (i.e. non-clearing markets), demand-led growth and supply constraints, integrating the Input–Output Analysis of 35 industrial sectors and households.

2) **Energy**: This module includes the renewable and non-renewable energy resources potentials and availability considering biophysical and temporal constraints. In total, 5 final fuels are considered (electricity, heat, solids, gases and liquids) and a diversity of energy technologies are modelled, following a net energy approach.

3) **Infrastructures**: Energy infrastructures represent the power plants to generate electricity and heat.

4) **Materials**: Materials are required by the economy and MEDEAS tracks the material requirements for the construction and the Operations and Maintenance of the infrastructures.

5) **Land Use**: It mainly accounts for the land requirements of the RES.

6) **Climate Change**: This module projects the climate change levels due to the GHG emissions generated by the human societies, which also feed-back through a damage function.

7) **Social and Environmental Impacts Indicators**: this module translates the "biophysical" results of the simulations into metrics related with social and environmental impacts. The objective is to contextualize the implications for human societies in terms of well-being.

The modules of economy and energy are the most extensive and reach the highest degree of disaggregation. The modules have bi-directional linkages, excepting for the Land-use and Social and Environmental impacts indicators, which mainly report outputs from the simulations without feed-backing to rest of the structure (see Figure 1).

In this work, we apply an optimization algorithm to fit with MEDEAS model the GHG emission scenarios provided by IPCC and integrated by INSTM according to Global Warming 2 °C consistent. The aim is to explore the capability of the model in reproduce experimental data (scenarios), i.e., to search the optimized parameters in order to have some indications how to reach the global warming of 2 °C in practice.

Model optimization or identification is a common mathematical technique to calculate the parameters of a dynamical system in order to fit experimental data that represent a physical phenomenon with its model representation. In literature, we find several examples, in mechanics, engineering, economics and finance, geophysics, biology, ecology (Dorfman, 1969; Marsili-Libelli, 1992; Martelloni et al., 2013; Santarlasci et al., 2014).

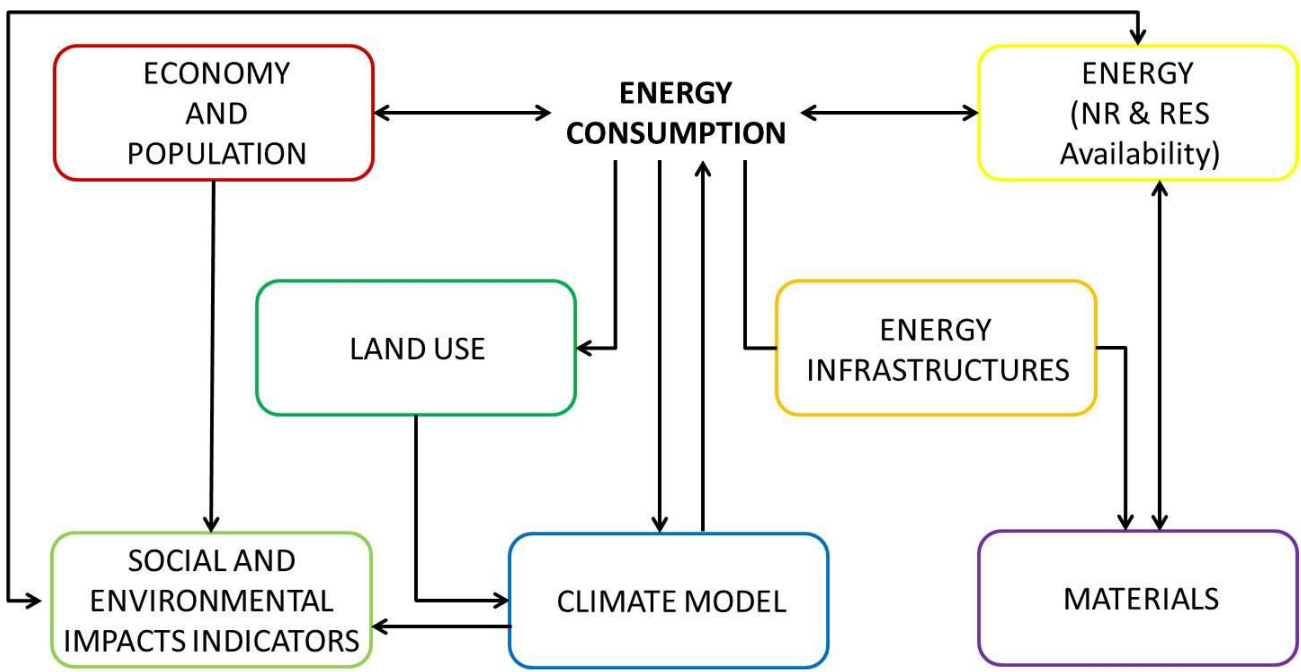

Figure 1. Overview of MEDEAS-World by modules and the linkages between them (https://medeas.eu/model/medeas-model).

**MATERIALS AND METHOD**

The Optimization methodology is implemented in MATLAB®, which allows running the Python platform in which MEDEAS is developed.

In this context, we do not report all the details of the optimization procedure that is schematically described in the block diagram of Figure 2.

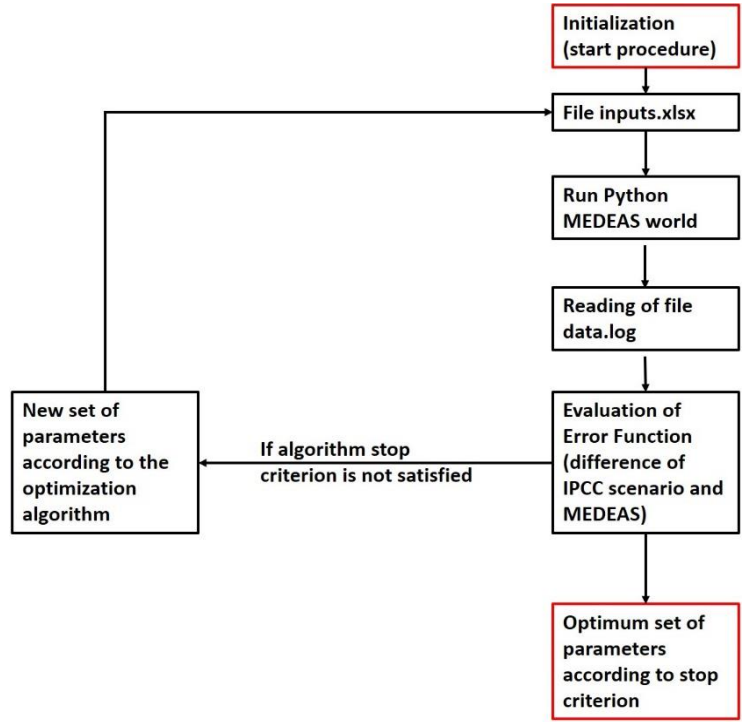

Figure 2. Block diagram exploiting the optimization procedure.

Here we report the results of the optimization procedure obtained using SIMPSA algorithm (Cardoso, 1996) to fit IPCC emission scenario at world level with starting policy from year 2020 in order to reduce GHG emissions (increment of RES and reduction of fossil fuels). This optimization algorithm finds the best set of parameters P = [k1, ..., ki, ..., kn] which minimize error function F(P),

$$F(\mathbf{P}) = \frac{1}{N}\sum_{i=1}^{N}\left(x_i^{exp} - x_i^{mod}(\mathbf{P})\right)^2, \qquad (1)$$

where $x^{exp}$ and $x^{mod}$ indicate respectively the "experimental" (IPCC/INSTM scenario) and the values obtained by the MEDEAS model on time (i represents the number of data relative to time in years). The optimization procedure stops when the value of the error function is lower than an arranged threshold (see Figure 2).

MEDEAS model was extremely complex as it contained many parameters and variables (about 4000), so the strategy was to focus on the calibration of the parameters that directly influenced the emissions as annual growth of RES parameters, the rates of coal and oil extraction, GDP pro-capite (GDPpc) and annual population growth (see Results section). We adjusted these parameters starting from Business As Usual (BAU) scenario which represents the Baseline scenario. In the BAU scenario, no new policies or measures to reduce GHG emissions have been implemented, apart from those already adopted. Therefore, varying the parameters in Table 1, we obtained several new scenarios which differed from the BAU in term of emission pathways. In this analysis, it is central to set the search ranges in which the parameters vary, in order to obtain from calibration a set with physical sense. For a complete description of the MEDEAS parameters, see Capellán-Pérez et al. (2017a).

**RESULTS**
We show the result of the first optimization in which we consider 39 parameters, three for GDPpc, two for the population growth and 34 other ones as indicated in Table 1 where we specify the initial values for the calibration. For GDPpc we set three ranges for the calibration procedure: [0.02 0.065] from 2015 to 2020, [0.005 0.015] from 2021 to 2025 and [-0.025 -0.001] from 2026 to 2050; the GDPpc initialization values are respectively 0.06, 0.01 and -0.015. By calibration we obtain respectively for the GDP intervals the values 0.0612, 0.01 and -0.0149. While for the annual population growth we use a linear function in which the slope is imposed to vary from -0.00041 to -0.00039 and the intercept from 0.8 to 0.82 (see in Figure 3 the result from calibration); the initialization values are respectively -0.0004 and 0.8098 (the optimized values are respectively -0.00039986 and 0. 80986879).

The choice of the initialization values is due to the simple reasons to approach the reference curve reported in figure 5, i.e., the IPCC/INSTM scenario (red line). This curve has been elaborated considering the global remaining carbon budget estimation in the Climate Action Tracker (CAT) "2°C consistent scenario", which concerns a warming below 2 °C with at least a 66% probability over the whole of the 21st century. The red curve follows the trend from IPCC AR5 scenarios (from Working Group III) from 2012 up to 2020, then it declines to assure the same carbon budget of 1494 GtCO2eq delimited by the 2°C consistent scenario. AR 5 scenario considers, in absence of policies global warming, to reach 4.1 °C – 4.8 °C above pre-industrial by the end of the century (higher and lower trend assessments with no policy, black lines Figure 5) with a carbon budget much higher than the 2°C median considered here the desirable target.

The CAT database also includes other scenarios, in particular we considered the following groups, to be compared with the reference elaborated by IPCC/INSTM (red line):
- 1.5 °C consistent scenarios (high, median, low; pink lines)
- future emissions under current policies through 2030. The scenarios cover implemented policies at the time of the update, and other developments such as expected economic growth or trends in activity and energy consumption (figure 5, blue lines);
- pledge policies in 2020 are the result of Copenhagen pledges (unconditional and conditional pledges). The effort-sharing bar allows the evaluation of their level of adequacy (figure5, purple lines).

Regarding the other parameters involved in the optimization we set the ranges as indicated in Table 2. In Figure 4 we show the result of the fitting in which we can note the according between the two emission curves expressed in GT (Gigaton).

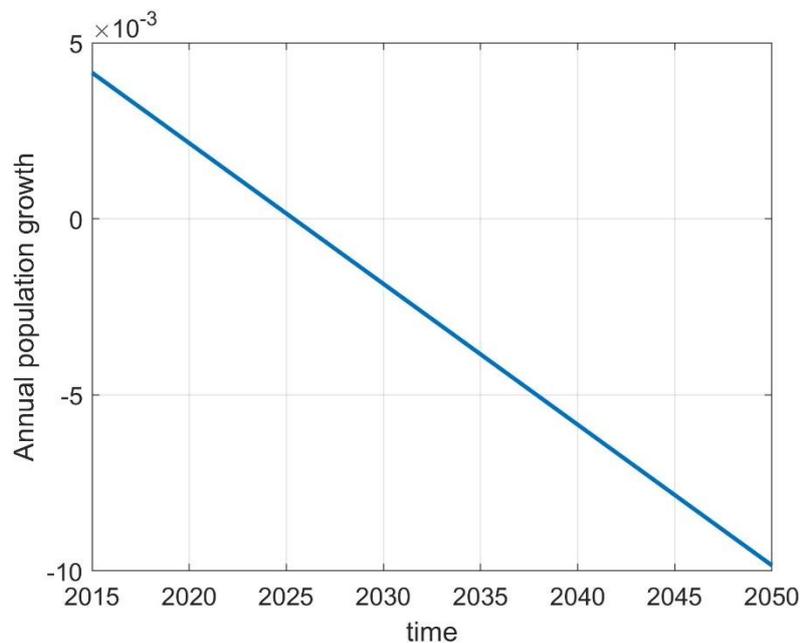

Figure 3. Annual population growth obtained by means of optimization algorithm.

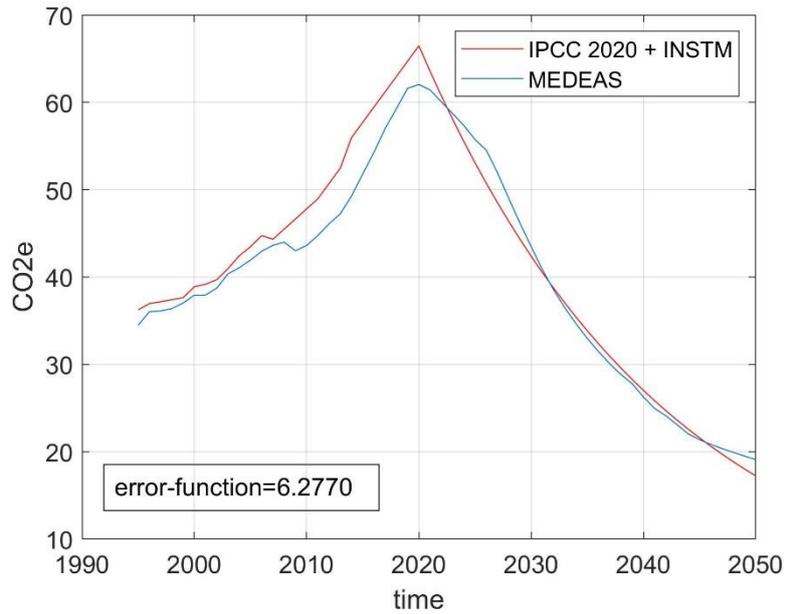

Figure 4. Total CO$_2$ equivalent (all GHGs) obtained by SIMPSA optimization with 39 calibrated parameters.

Table 1. Initial values of parameters used for the calibration.

| Parameters | Initial values | Parameters | Initial values |
|---|---|---|---|
| hydro growth | 2,8% | solar for heat | 10% |
| geot-elec growth | 2,4% | geothermal for heat | 7,7% |
| solid bioE-elec growth | 7,2% | solid bioE for heat | 6,3% |
| oceanic | 10% | Policy electric household 4wheeler vehicle Tfin | 0,0064 |
| onshore wind | 15% | Policy hybrid household 4w vehicle Tfin | 0,0108 |
| wind offshore | 15% | Policy gas household vehicle 4w Tfin | 0,1489 |
| solar PV (Photovoltaic) | 15% | Policy electric 2wheeler h. Tfin | 0,9254 |
| biofuels 2gen | 5% | Policy change to 2wheeler h. Tfin | 0,3325 |
| biofuels 3gen | 5% | Policy hybrid HV (heavy vehicles) Tfin | 0,00045 |
| bioE residues for heat+elec | 5% | Policy gas HV Tfin | 0,00045 |
| cellulosic biofuels | 8% | Policy electric LV (light cargo vehicles) Tfin | 0,00074 |
| waste change | 0,04436 | Policy hybrid LV Tfin | 0,00036 |
| BEV (Battery Electric Vehicle) growth | 9% | Policy gas LV Tfin | 0,01597 |
| HEV (Hybrid Electric Vehicle) growth | 9% | Policy electric bus Tfin | 0 |
| NGV (Natural Gas Vehicle) growth | 7% | Policy hybrid bus Tfin | 0 |
| PHS (Pumped Hydro Storage) | 2,3% | Policy gas bus Tfin | 0 |
| CSP (Concentrated Solar Power) | 10% | Policy electric train Tfin | 0,2 |

Table 2. All parameters regarding optimization shown in Figure 4, we report the optimized values in column 2 and the set ranges for the calibration in column 3.

| Parameters | Optimized values (Figure 10) | Range for the optimization |
|---|---|---|
| P hydro growth | 30,3% | 0 - 50 % |
| P geot-elec growth | 29,2% | 0 - 50 % |
| P solid bioE-elec growth | 17,4% | 0 - 50 % |
| P oceanic | 24% | 0 - 50 % |
| P onshore wind | 21% | 0 - 50 % |
| P wind offshore | 9% | 0 - 50 % |
| P solar PV (Photovoltaic) | 45% | 0 - 50 % |
| P biofuels 2gen | 15,0% | 0 - 50 % |
| P biofuels 3gen | 27,0% | 0 - 50 % |
| P bioE residues for heat+elec | 16,0% | 0 - 50 % |
| P cellulosic biofuels | 33% | 0 - 50 % |
| P waste change | 0,025465277 | 0 - 0,1 |
| P BEV (Battery Electric Vehicle) growth | 28% | 0 - 50 % |
| P HEV (Hybrid Electric Vehicle) growth | 34% | 0 - 50 % |
| P NGV (Natural Gas Vehicle) growth | 8% | 0 - 50 % |
| P PHS (Pumped Hydro Storage) | 20,0% | 0 - 50 % |
| P CSP (Concentrated Solar Power) | 41% | 0 - 50 % |
| P solar for heat | 29,6% | 0 - 50 % |
| P geothermal for heat | 32,2% | 0 - 50 % |
| P solid bioE for heat | 14,3% | 0 - 50 % |
| Policy electric household 4wheeler vehicle Tfin | 0,52420319 | 0 - 1 |
| Policy hybrid household 4w vehicle Tfin | 0,309401608 | 0 - 1 |
| Policy gas household vehicle 4w Tfin | 0,031106111 | 0 - 1 |
| Policy electric 2wheeler h. Tfin | 0,655017852 | 0 - 1 |
| Policy change to 2wheeler h. Tfin | 0,597248823 | 0 - 1 |
| Policy hybrid HV Tfin | 0,164336529 | 0 - 1 |
| Policy gas HV Tfin | 0,517193803 | 0 - 1 |
| Policy electric LV Tfin | 0,243356834 | 0 - 1 |
| Policy hybrid LV Tfin | 0,081390031 | 0 - 1 |
| Policy gas LV Tfin | 0,496621902 | 0 - 1 |
| Policy electric bus Tfin | 0,200234358 | 0 - 1 |
| Policy hybrid bus Tfin | 0,680892168 | 0 - 1 |
| Policy gas bus Tfin | 0,011117541 | 0 - 1 |
| Policy electric train Tfin | 0,887146369 | 0 - 1 |

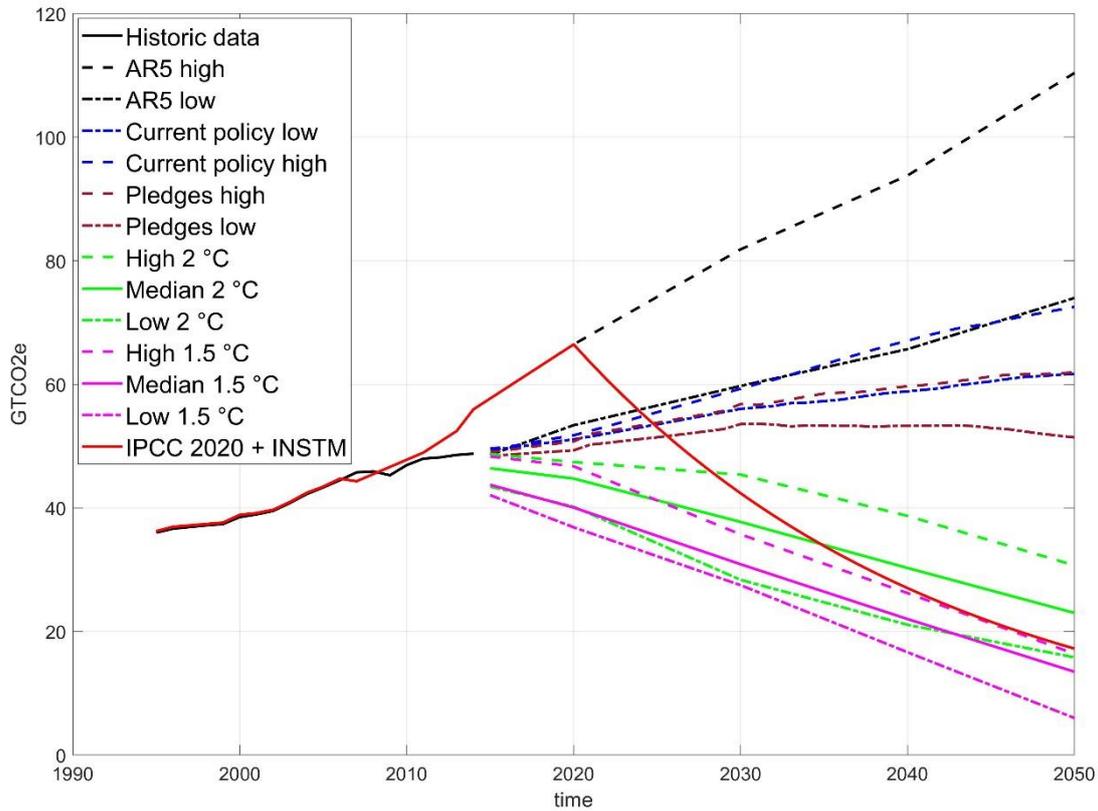

Figure 5. The red line is the IPCC2020 + INSTM scenario, the blue line is the MEDEAS simulation of Figure 4, but considering GDP of Table 2; the dark dash and dark dash-dot are respectively the AR5 BAU high and low; the blue dash and dark dash-dot are respectively the current policy high and low; the purple dash and dark dash-dot are respectively the pledge policy high and low; the green and green dash are respectively the median and high warming projection 2 °C consistent; the blue dash is the optimized MEDEAS simulation according to projection 2 °C consistent (high); the dark green is warming projection 1.5 °C consistent, reported for comparison with MEDEAS simulations.

After to perform the above preliminary results, we achieve other simulations in order to calibrate the warming projection 2 °C consistent (high, median and low) provided by IPCC. First, we initialize the RES parameters considering the values indicated in Table 3 and the transportation policy ones (Table 4). Then to initialize annual population and GDPpc growth we consider the available data by world data bank (https://data.worldbank.org) and we fit both by means of a linear function (see Figures 6 and 7) in order to have 4 parameters. Moreover, by means of the obtained laws we forecast the GDP and annual population growth up to 2050 to complete the input file for simulation of the MEDEAS model. Therefore, in these calibrations, we consider 38 parameters, i.e., the RES ones and 4 parameters (slopes and intercepts) for the GDPpc and annual population growth.

Table 1. Initial values of RES parameters used for the calibration of scenarios high, median and low warming projection 2 °C consistent.

| Parameters | High | Median | Low |
|---|---|---|---|
| hydro growth | 10% | 20% | 25% |
| geot-elec growth | 4,8% | 4,8% | 4,8% |
| solid bioE-elec growth | 14,4% | 14,4% | 14,4% |
| oceanic | 15% | 15% | 15% |
| onshore wind | 20% | 20% | 20% |
| wind offshore | 20% | 20% | 20% |
| solar PV (Photovoltaic) | 8% | 8% | 8% |
| biofuels 2gen | 8% | 8% | 8% |
| biofuels 3gen | 8% | 8% | 8% |
| bioE residues for heat+elec | 8% | 8% | 8% |
| cellulosic biofuels | 8% | 8% | 8% |
| waste change | 0,04436 | 0,04436 | 0,04436 |
| BEV (Battery Electric Vehicle) growth<br>HEV (Hybrid Electric Vehicle) growth<br>NGV (Natural Gas Vehicle) growth | 10%<br>10%<br>5% | 20%<br>20%<br>5% | 25%<br>25%<br>10% |
| PHS (Pumped Hydro Storage)<br>CSP (Concentrated Solar Power) | 10%<br>10% | 20%<br>20% | 25%<br>25% |
| solar for heat | 10% | 20% | 25% |
| geothermal for heat | 10% | 20% | 25% |
| solid bioE for heat | 10% | 20% | 25% |

Table 2. Initial values of transportation policy parameters used for the calibration of scenarios high, median and low warming projection 2 °C consistent.

| Parameters | High | Median | Low |
|---|---|---|---|
| Policy electric household 4wheeler vehicle Tfin<br>Policy hybrid household 4w vehicle Tfin<br>Policy gas household vehicle 4w Tfin<br>Policy electric 2wheeler h. Tfin<br>Policy change to 2wheeler h. Tfin | 0,38<br>0,25<br>0,15<br>0,95<br>0,6 | 0,38<br>0,25<br>0,15<br>0,95<br>0,6 | 0,45<br>0,25<br>0,15<br>0,98<br>0,6 |
| Policy hybrid HV Tfin | 0,7 | 0,7 | 0,75 |
| Policy gas HV Tfin | 0,15 | 0,15 | 0,15 |
| Policy electric LV Tfin | 0,4 | 0,4 | 0,45 |
| Policy hybrid LV Tfin | 0,2 | 0,2 | 0,2 |
| Policy gas LV Tfin | 0,15 | 0,15 | 0,15 |
| Policy electric bus Tfin | 0,4 | 0,4 | 0,42 |
| Policy hybrid bus Tfin | 0,4 | 0,4 | 0,4 |
| Policy gas bus Tfin | 0,08 | 0,08 | 0,08 |
| Policy electric train Tfin | 0,8 | 0,8 | 0,85 |

In Figures 8-10 we report respectively the results of obtained calibration of the three emission scenarios (high, median and low) by means of MEDEAS model and the related (optimized) annual population and GDPpc growth. In Tables 5 and 6 we report the related optimized parameters for all calibrations. We observe that the RES parameters, on average, increase from high to low warming 2 °C consistent, as we could be expected, but median are (on average) a little smaller of high. Moreover, we note that while the GDP decreases from high to law warming calibration, the annual growth increases, but this behavior probably depends from the shape of error functions in which the global minimum corresponds to high values of the annual population growth and low values of the GDPpc when we consider the three mentioned optimizations. In conclusion, we can assert that GDPpc and annual population growth influence the model response more importantly than RES parameters.

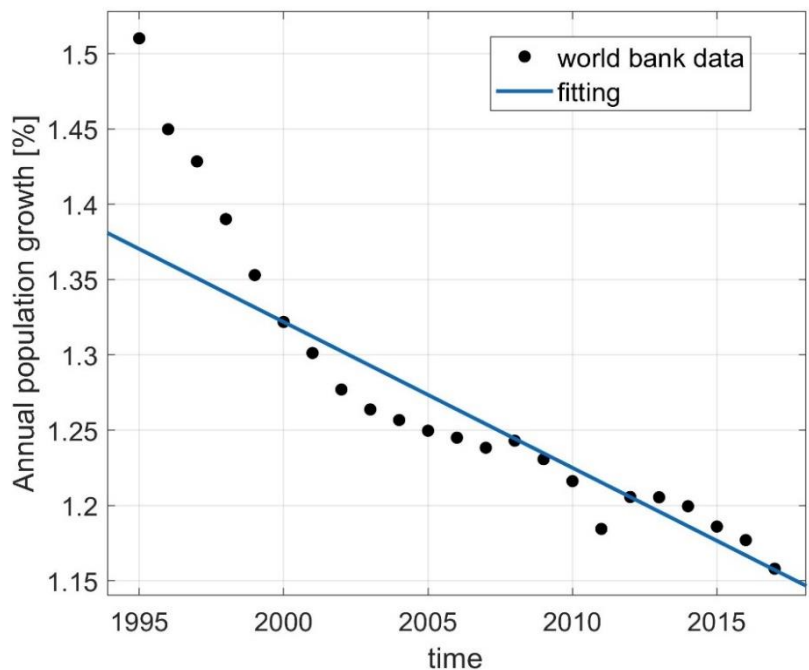

Figure 6. Fit of the annual population growth from 1995 to 2017 (data extracted from the world bank data).

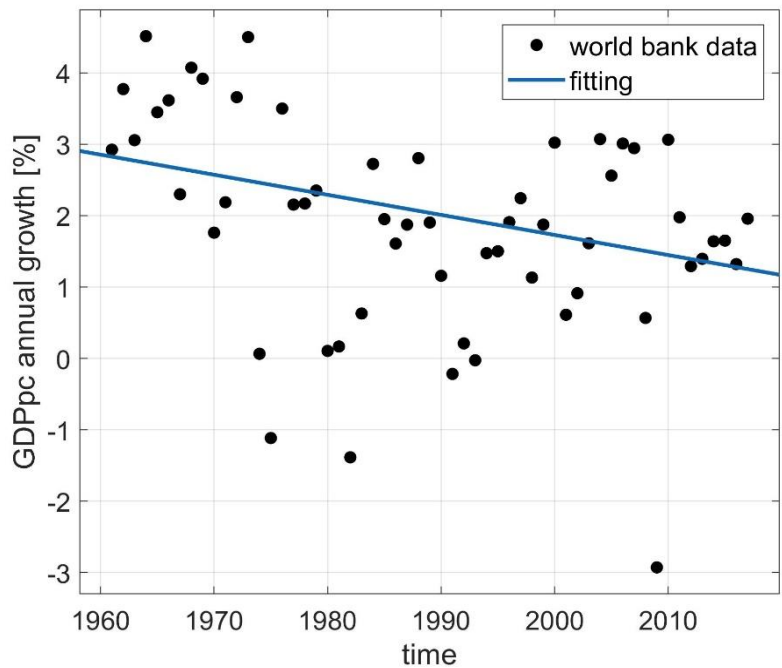

Figure 7. Fit of the GDP pro capite annual growth from 1961 to 2017 (data extracted from the world bank data).

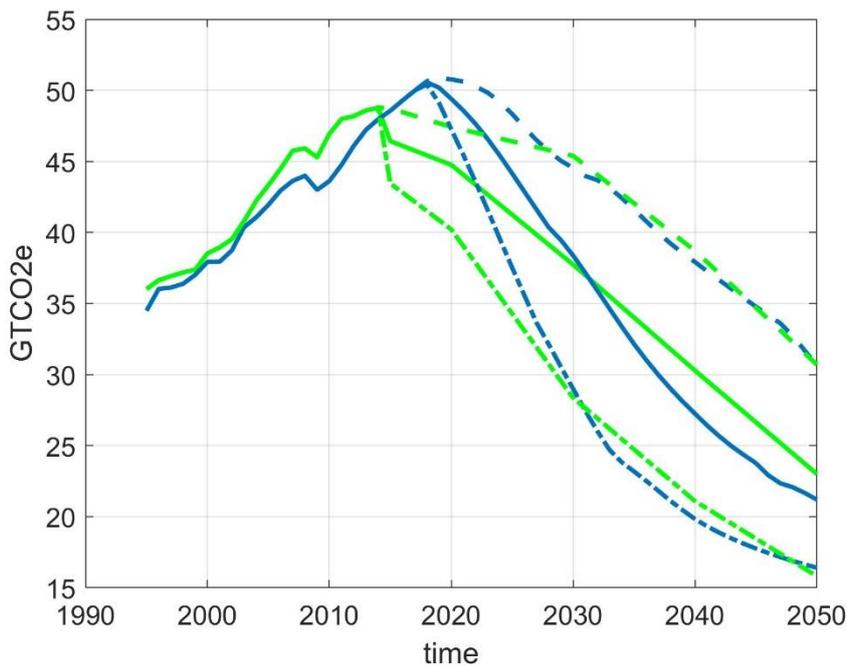

Figure 8. Optimization results: the green curves and blue ones are respectively high IPCC warming projection 2 °C consistent (dash green), MEDEAS high optimization (dash light blue), median IPCC warming projection 2 °C consistent (solid green), MEDEAS median optimization (solid light blue), low IPCC warming projection 2 °C consistent (dash-dot green), MEDEAS low optimization (dash-dot light blue). The values of error-functions are respectively (from high to low) 2.423, 6.55 and 9.66.

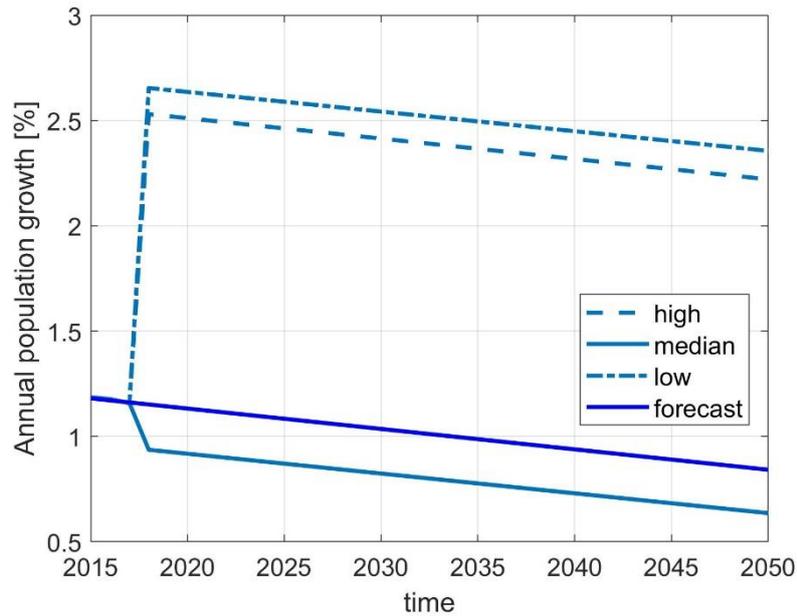

Figure 9. The calibrated annual population growth: the solid light blue corresponds to the forecast up to 2050 (starting from data of World Data Bank) in order to initialize the calibration procedure, the dash-dot blue is the result from calibration according to the high IPCC warming projection 2 °C consistent, the solid blue is the result from calibration according to the median IPCC warming projection 2 °C consistent and the dash blue is the result from calibration according to the low IPCC warming projection 2 °C consistent.

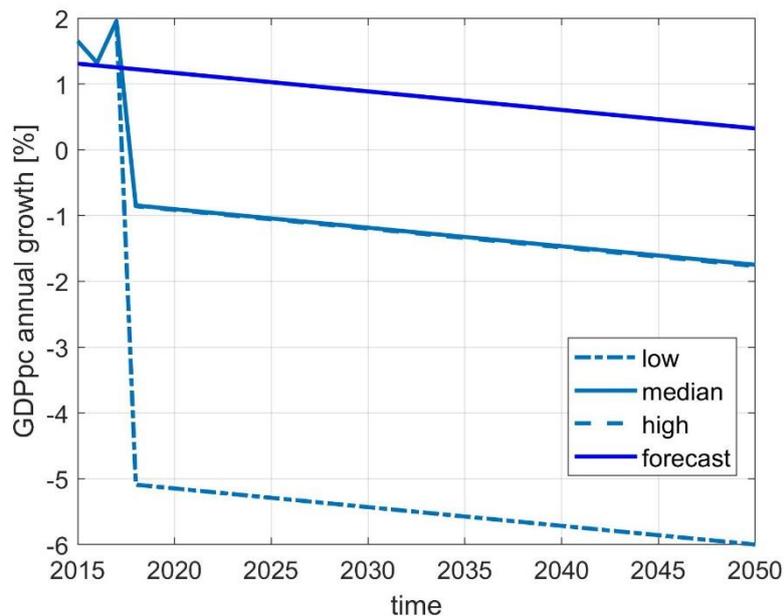

Figure 10. The calibrated GDPpc annual growth: the solid light blue corresponds to the forecast up to 2050 (starting from data of World Data Bank) in order to initialize the calibration procedure, the dash-dot blue is the result from calibration according to the high IPCC warming projection 2 °C consistent, the solid blue is the result from calibration according to the median IPCC warming projection 2 °C consistent and the dash blue is the result from calibration according to the low IPCC warming projection 2 °C consistent.

Table 5. Obtained RES parameters from the calibration scenarios high, median and low warming projection 2 °C consistent.

| Parameters | Optimized values IPCC High | Optimized values IPCC Median | Optimized values IPCC Low | Fixed ranges for the optimization |
|---|---|---|---|---|
| hydro growth | 13,2% | 18,9% | 21,2% | 0 - 40 % |
| geot-elec growth | 29,9% | 6,4% | 1,3% | 0 - 40 % |
| solid bioE-elec growth | 9,5% | 6,5% | 10,1% | 0 - 40 % |
| oceanic | 5% | 10% | 11% | 0 - 40 % |
| onshore wind | 18% | 12% | 26% | 0 - 40 % |
| wind offshore | 11% | 14% | 14% | 0 - 40 % |
| solar PV (Photovoltaic) | 33% | 30% | 34% | 0 - 40 % |
| biofuels 2gen | 9% | 1% | 14% | 0 - 40 % |
| biofuels 3gen | 16% | 23% | 19% | 0 - 40 % |
| bioE residues for heat+elec | 18% | 8% | 5% | 0 - 40 % |
| cellulosic biofuels | 22% | 16% | 8% | 0 - 40 % |
| waste change | 0,0375 | 0,087 | 0,08 | 0 - 0,1 |
| BEV (Battery Electric Vehicle) growth | 15% | 20% | 26% | 0 - 40 % |
| HEV (Hybrid Electric Vehicle) growth | 25% | 24% | 26% | 0 - 40 % |
| NGV (Natural Gas Vehicle) growth | 8% | 5% | 7% | 0 - 40 % |
| PHS (Pumped Hydro Storage) | 12,1% | 13,7% | 24% | 0 - 40 % |
| CSP (Concentrated Solar Power) | 19% | 24% | 25% | 0 - 40 % |
| solar for heat | 37,3% | 20 % | 16,4% | 0 - 40 % |
| geothermal for heat | 4,5% | 22,3 % | 12,8% | 0 - 40 % |
| solid bioE for heat | 11,6% | 12,4% | 24,7% | 0 - 40 % |

Table 6. Obtained transportation policy parameters from the calibration scenarios high, median and low warming projection 2 °C consistent.

| Parameters | Optimized values IPCC High | Optimized values IPCC Median | Optimized values IPCC Low | Ranges |
|---|---|---|---|---|
| Policy electric household 4wheeler vehicle Tfin | 0,392762349 | 0,358323364 | 0,43622925 | 0,25–0,5 |
| Policy hybrid household 4w vehicle Tfin | 0,266355049 | 0,237260522 | 0,235172001 | 0,2-0,3 |
| Policy gas household vehicle 4w Tfin | 0,151670479 | 0,149748334 | 0,144501054 | 0,1-0,2 |
| Policy electric 2wheeler h. Tfin | 0,976435414 | 0,960548833 | 0,970416746 | 0,9-1 |
| Policy change to 2wheeler h. Tfin | 0,589747363 | 0,612476454 | 0,593670363 | 0,5-0,7 |
| Policy hybrid HV Tfin | 0,706706735 | 0,677830554 | 0,744115351 | 0,5-0,8 |
| Policy gas HV Tfin | 0,151037035 | 0,127825973 | 0,161442129 | 0,1-0,2 |
| Policy electric LV Tfin | 0,428289168 | 0,35085463 | 0,468888008 | 0,3-0,5 |
| Policy hybrid LV Tfin | 0,193439397 | 0,209949917 | 0,168274521 | 0,15-0,3 |
| Policy gas LV Tfin | 0,194992338 | 0,167638513 | 0,166511669 | 0,1-0,2 |
| Policy electric bus Tfin | 0,409098992 | 0,413108197 | 0,424847501 | 0,3-0,45 |
| Policy hybrid bus Tfin | 0,413501323 | 0,401079742 | 0,428476466 | 0,3-0,45 |
| Policy gas bus Tfin | 0,078286012 | 0,069260505 | 0,082113826 | 0,05-0,1 |
| Policy electric train Tfin | 0,863405753 | 0,798483824 | 0,813145664 | 0,7-0,9 |

**SENSITIVITY ANALYSIS**

In this section we show the results related to sensitivity analysis. In Table 7 we report the RES parameters involved in the sensitivity analysis shown in Figure 11. We observe that, starting to the value of GDPpc and annual population growth according to median warming 2 °C consistent (obtained from model output), varying concurrently the RES parameters from 0 to 50% (step equal to 2 %), it is not possible to reach the low configuration. We note that we do not consider, in this analysis, the parameters involved in the transportation as we test the less influence (almost negligible) than other parameters on GHG emission. Indeed, we increment the policy of the electric vehicles (imposed to the maximum value equal to one) without obtaining important changes (see the dot blue curve in Figure 11). While, considering again the median model configuration but varying the GDPpc (sensitivity analysis related to this parameter), it is possible to reach the low configuration (see Figure 12). In Figure 13 we report the corresponding values of GDPpc that we vary in order to obtain a range that contains the optimized GDPpc curves. In Figures 14 and 15 we report the sensitivity results on the annual population growth applying the same methodology for GDPpc, i.e., we fix all the parameters obtained from median optimization and we vary only the annual population growth as reported in Figure 15. We note that the results of Figures 12 and 14 are similar, but the ranges of GDPpc and annual population growth (Figures 13 and 15) are different. In particular, the range of annual population growth (to obtain the same results, varyng GDPpc, in term of GHG emissions) is larger than GDPpc one compared to the same optimized curves (see Figures 13 and 15).

Table 7. The parameters (RES annual growth) involved in the sensitivity analysis: we consider for each simulation an increment of 2% for all parameters with exclusion of waste change parameter that vary from 0 to 0.1 with step equal to 0.004.

| Parameters | Δ=2% |
|---|---|
| hydro growth | 0 - 50 % |
| geot-elec growth | 0 - 50 % |
| solid bioE-elec growth | 0 - 50 % |
| oceanic | 0 - 50 % |
| onshore wind | 0 - 50 % |
| wind offshore | 0 - 50 % |
| solar PV (Photovoltaic) | 0 - 50 % |
| biofuels 2gen | 0 - 50 % |
| biofuels 3gen | 0 - 50 % |
| bioE residues for heat+elec | 0 - 50 % |
| cellulosic biofuels | 0 - 50 % |
| waste change | 0 – 0,1 |
| BEV (Battery Electric Vehicle) growth | 0 - 50 % |
| HEV (Hybrid Electric Vehicle) growth | 0 - 50 % |
| NGV (Natural Gas Vehicle) growth | 0 - 50 % |
| PHS (Pumped Hydro Storage) | 0 - 50 % |
| CSP (Concentrated Solar Power) | 0 - 50 % |
| solar for heat | 0 - 50 % |
| geothermal for heat | 0 - 50 % |
| solid bioE for heat | 0 - 50 % |

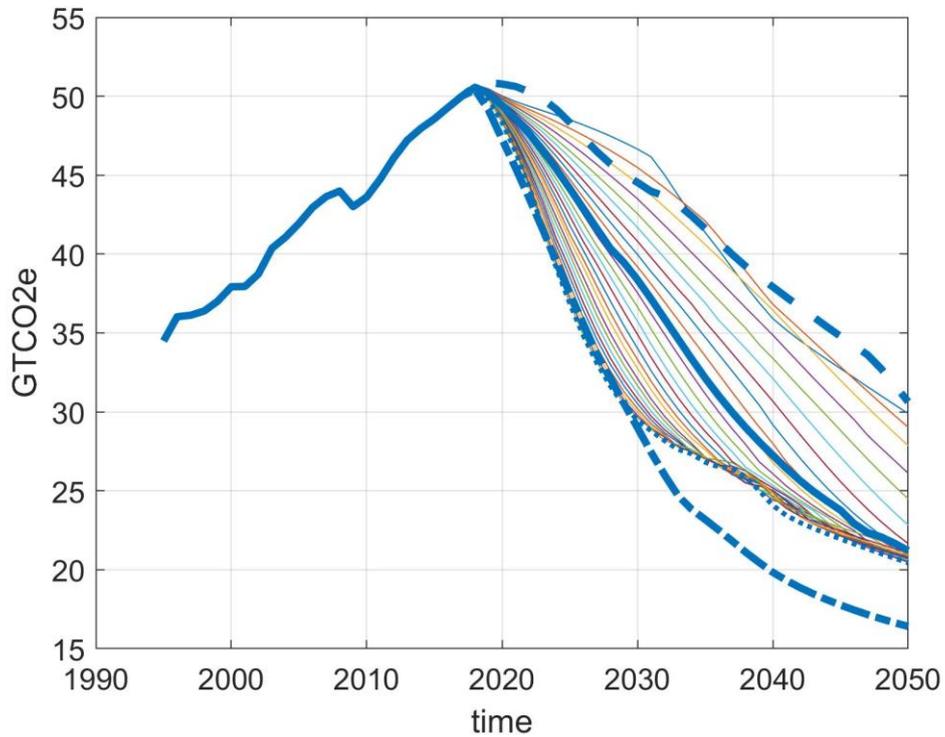

Figure 11. Sensitivity analysis according to RES parameters reported in Table 3. The blue dash, solid and dash-dot correspond to calibrated output of the MEDEAS model according respectively with high, median and low warming 2 °C consistent; the dot curve represents the last one with 50% RES parameter values and maximizing the transportation parameters (the obtained gain is relatively small).

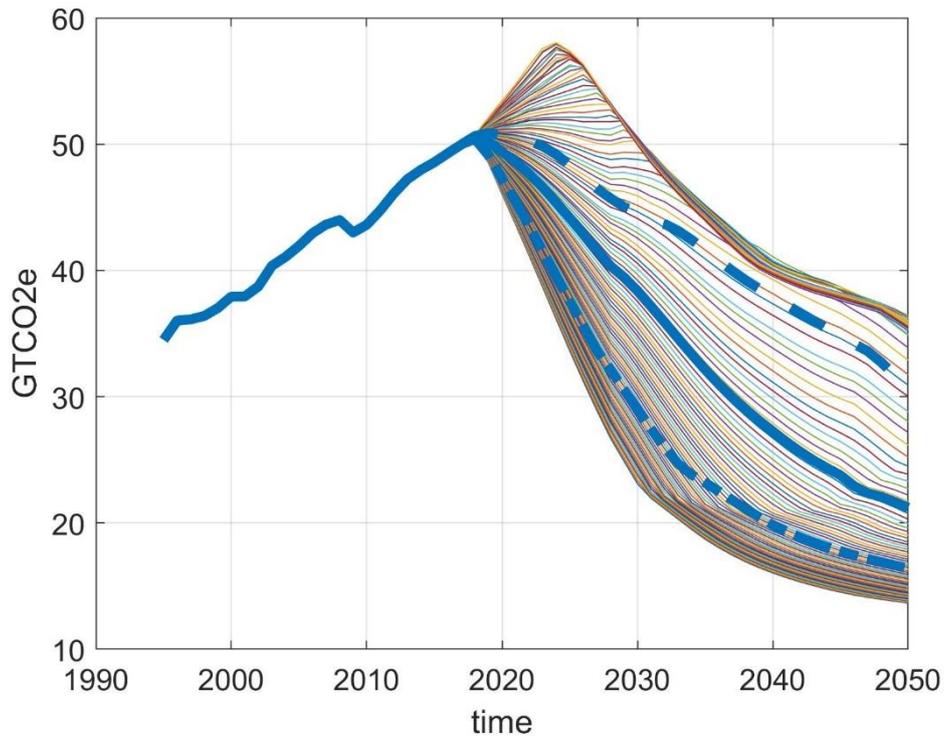

Figure 12. Sensitivity analysis according to GDPpc variations reported in Figure 13. The blue dash, solid and dash-dot correspond to calibrated output of the MEDEAS model according respectively with high, median and low warming 2 °C consistent.

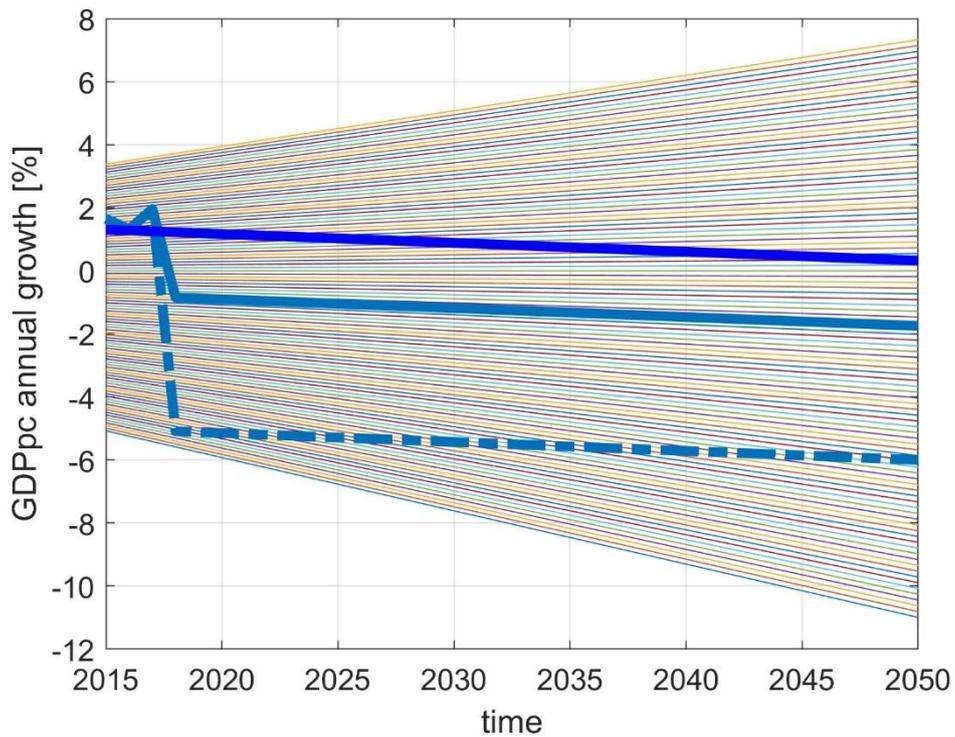

Figure 13. Considered GDPpc variations in order to study the sensitivity analysis of the MEDEAS model (see Figure 12). We report also the calibrated GDPpc annual growth as in Figure 10.

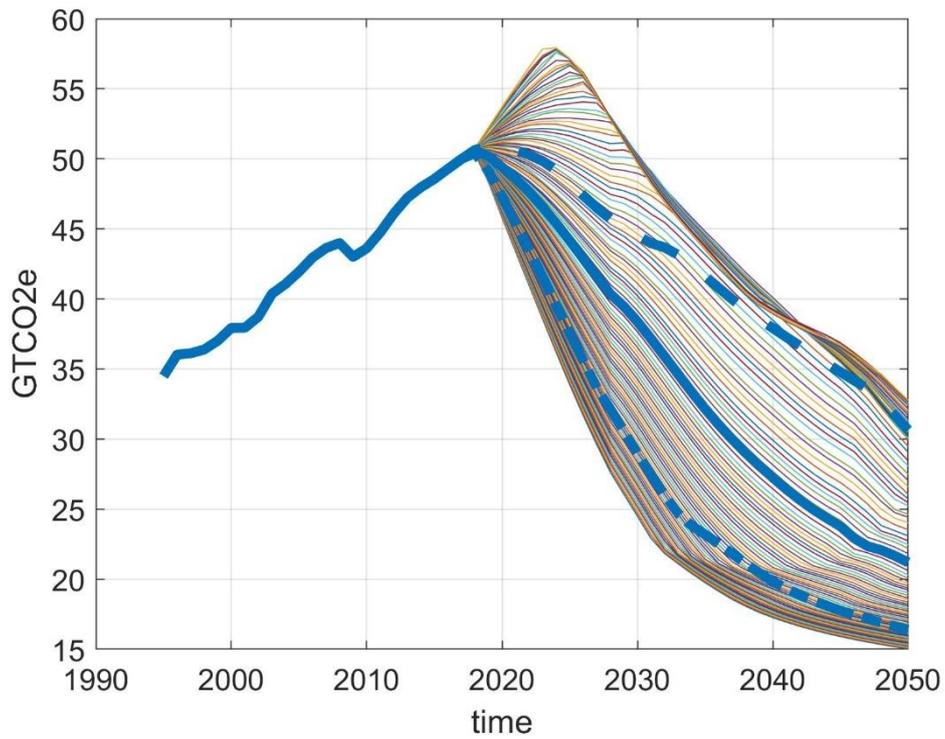

Figure 14. Sensitivity analysis according to annual population growth variations reported in Figure 15. The blue dash, solid and dash-dot correspond to calibrated output of the MEDEAS model according respectively with high, median and low warming 2 °C consistent.

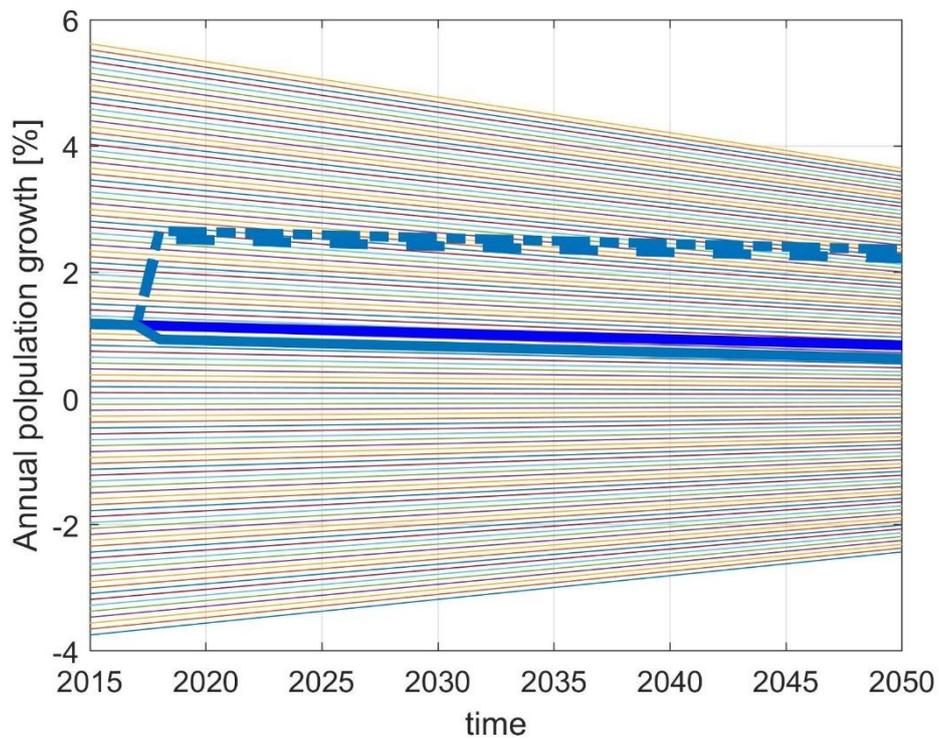

Figure 15. Considered annual population growth variations in order to study the sensitivity analysis of the MEDEAS model (see Figure 14). We report also the calibrated annual population growth as in Figure 9.

**CONCLUSIONS AND DISCUSSION**

A new methodology for building scenarios trying to fit 2020 decarbonization scenarios have been proposed for the Python Version of MEDEAS world (pmedeas_w) using optimization algorithms developed in the MATLAB® environment. The methodology is still under development, anyway preliminary results show the promising use of pymedeas_w interface with other platforms, especially in the view to use MEDEAS not only as models to explore simulated scenarios but also as models to fit experimental data. However, to reproduce theoretical scenarios in a large parameter domain, it is necessary to investigate the reliability of the values of the involved parameters, first of all GDP and RES growth, to which the model presents a certain sensibility.

In particular the sensibility of GDP variable is due to the MEDEAS model approach in linking the economy and the energy modules, connected to provide dynamic adaptive scenarios. In most of the Integrated Assessment Models (IAMs), used for instance, by IPCC, energy supply is no subject to limits, thus, the results underestimate the real effects of resources' scarcity on GDP. Thanks to the MEDEAS model peculiarity, that provides feedbacks between economy and energy availability, it is possible to cover the lack of consideration of energy and resources scarcity on the economy, and, in addition to the initial conditions imposed by the exogenous scenarios framework, the results are endogenously adapted to account for energy and resources physical limits.

Acknowledgements:

We acknowledge the support of MEDEAS project through the funding from the European Union's Horizon 2020 research and innovation programme under grant agreement No 691287.